\documentclass{aa}
\usepackage{graphics}
\usepackage{txfonts}

\begin{document}

\def\te{T_{\rm eff} }
\def\delt{\scriptstyle \Delta}

\def\eol{\hfil\break}
\def\va{\vskip 5.0 mm} \def\vb{\vskip 2.5 mm} \def\vc{\vskip 1.5 mm}
\def\ha{\hskip 5.0 mm} \def\hb{\hskip 2.5 mm} \def\hc{\hskip 1.5 mm}
\def\vd{\vskip 1.0 mm} \def\hd{\hskip 1.0 mm}

   \title{Mass distribution of DA white dwarfs in the First Data Release
          of the Sloan Digital Sky Survey}

   \author{J. Madej\inst{1}, M. Nale\.zyty\inst{1},
           \and
           L.G. Althaus\inst{2} }

   \offprints{M. Nale\.zyty, \\ \email{nalezyty@astrouw.edu.pl}}
   \institute {$^1$Astronomical Observatory, University of Warsaw,
     Al. Ujazdowskie 4, 00-478 Warsaw, Poland \\
     $^2$ Departament de Fisica Aplicada, Universitat Politecnica de 
     Catalunya, Av. del Canal Olimpic s/n, 08860, Castelldefels, Spain.
   }

   \date{Received ...; accepted ...}

   \titlerunning{Mass distribution of DA white dwarfs in the SDSS DR1}

   \abstract{
     We investigate the sample of 1175 new nonmagnetic DA 
     white dwarfs with the effective temperatures $\te \ge 12000$ K,
     which were extracted from the Data Release 1 of
     the Sloan Digital Sky Survey. We determined masses, radii, and 
     bolometric luminosities of stars in the sample. The above parameters
     were derived from the effective temperatures $\te$ and surface
     gravities $\log g$ published in the DR1, and the new theoretical $M-R$
     relations for carbon -- core and oxygen -- core white dwarfs. 
     Mass distribution of white dwarfs in this sample exhibits the peak at
     $M=0.562 M_\odot$ (carbon core stars),
     and the tail towards higher masses. Both the shape of the mass
     distribution function and the empirical mass -- radius relation are
     practically identical for white dwarfs with either pure carbon or
     pure oxygen cores.

}

   \maketitle

%
%________________________________________________________________

\section{Introduction}

The Sloan Digital Sky Survey is the very extensive and ambitious
ground--based research project, aimed at the determination of positions,
brightnesses, and acquiring of the optical spectra of millions of various
celestial bodies. The SDSS was described in detail on their web page,
{\rm http://www.sdss.org/}, and in York et al. (2000). While this project
was essentially designed
to study galaxies, quasars, and large-scale structures in the Universe,
its secondary targets are stars of various spectral types. We take
advantage of the SDSS data on white dwarf stars of various classes, which
recently were made publicly available as the subset of the Data Release 1.

Comprehensive analysis of the photometric and spectral data on white dwarfs,
which are available in the DR1, was presented in Kleinman et al. (2004).
They have found 2935 white dwarfs of various types and related stars,
and only few of them were spectroscopically identified previously. The
number of well identified white dwarfs in their sample approaches
the number of white dwarf stars known previously, see McCook \& Sion
(1999), and the newest version of their catalogue available at
{\rm http://www.astronomy.villanova.edu/WDCatalog/index.html/.

Kleinman et al. (2004) performed fitting of the observed spectra of 1833
nonmagnetic DA white dwarfs to the widely used theoretical grid of pure 
hydrogen model atmospheres by Detlev Koester (see Finley, Koester \& Basri
1997). They have obtained effective temperatures in the range
$7220\le \te \le 93855$ K. Surface gravities of stars in the sample
range } from low values corresponding to subdwarfs to the most massive
white dwarfs of the highest $\log g = 9.0$.

The referee pointed out, that the SDSS DR1 values of $\log g$
determined for stars with $\te < 12000$ K are systematically larger than
for hotter stars, and are larger than expected for white dwarfs, cf.
Fig. 9 in Kleinman et al. (2004). The explanation of this effect was not
found, however, they favor the possibility that cooler model atmospheres
exhibit systematic errors in the parametrisation of convection, or in the
fitting procedure. As the result, they do not claim that the high values of 
$\log g$ are real. To avoid distortion of the high-mass tail presented
in Fig.~1 we analyse a subset of the original SDSS DR1 sample which 
includes 1175 DA white dwarfs with $\te \ge 12000$ K.

The Koester's grid does not include spectra of models with
high gravities, $\log g > 9.0$. Therefore, available determinations of
$\log g$ also do not exceed 9.0, and this is an artificial bias which
influences the determination of $\log g$ for over 20 most 
massive DA white dwarfs in the whole SDSS DR1 sample.

One can note, that white dwarfs of the highest gravity exceeding 9.0 appear
among previously known nonmagnetic DA stars. Compilation of the existing mass
determinations by Nale\.zyty \& Madej (2004) lists 6 such white dwarfs, 
with the maximum $\log g = 9.12$. Also the most recent paper by Dahn
et al. (2004) reports the measurement of $\log g = 9.46$ in the
apparently most massive star LHS 4033. Inferred mass of this nonmagnetic
DA star is $M \approx 1.33 \, M_\odot$, and the corresponding small radius
$R=0.00368 \, R_\odot$, place this star close to the limiting Chandrasekhar
mass.

\section{Mass distribution of DA white dwarfs} 

Masses $M$ and radii $R$ of the 1833 DA white dwarfs in the whole
SDSS DR1 sample
were directly determined from the corresponding values of $\te$ and $\log g$.
This method was introduced by Bergeron, Saffer \& Liebert (1992), and is 
recommended for the analysis of large samples of stars. Resulting
spectroscopic masses and radii, and their errors, strongly depend on the
accuracy and reliability of the stellar atmosphere and the hydrogen line
formation theory. 

Given $\te$ and $\log g$ of a star, the mass $M$ and radius $R$ were 
computed from the combination of the relations: \vskip 2 mm

\noindent {\bf 1.} Definition of the surface gravity $g$
\begin{equation}
M = {R^2 \over G} \, 10^{\log g} \, .
\label{equ:gravity}
\end{equation}
{\bf 2.} Theoretical mass -- radius relation
\begin{equation}
M = M(R, \te ) \, .
\label{equ:mr}
\end{equation}

\begin{figure}
   \resizebox{\hsize}{7.0cm}{\rotatebox{0}{\includegraphics{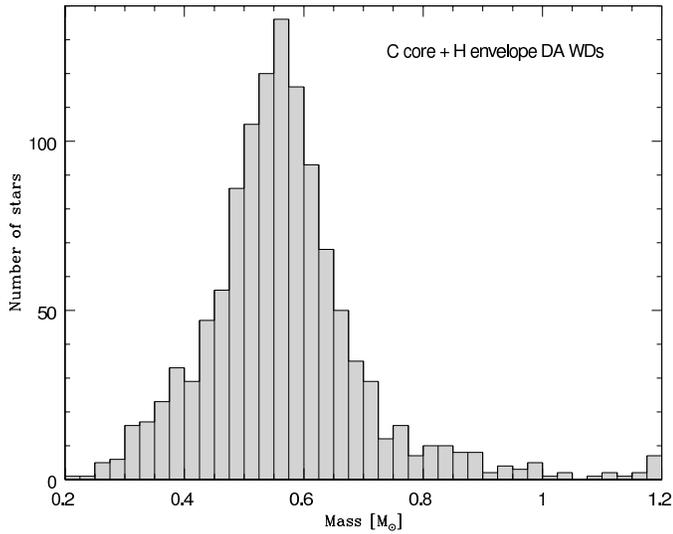}}}
   \caption[]{Mass distribution of the 1175 DA white dwarfs of the SDSS DR1 
      sample with $\te \ge 12000$ K. The peak mass is located at
      $M = 0.562 M_\odot$. Individual
      stars were grouped in bins of $0.025 M_\odot$ width.      }
   \label{fig:fig1}
\end{figure}

We have used here $M-R$ relations by Panei et al. (2000), which are briefly
discussed in the following Section. The reference set of spectroscopic
$M$ and $R$ values was obtained assuming $M-R$ relations for pure carbon
core and hydrogen envelope of all white dwarfs. 
Assumed mass of the envelope, $M_{\rm H}/M = 10^{-5}$.

Computational results of our research are displayed in Figs. 1$-$2. 
Fig. 1 presents the mass distribution of 1175 DA white dwarfs 
with $\te \ge 12000$ K.
Stars are grouped in relatively narrow bins of 0.025 $M_\odot$ width.
Masses of investigated stars range from $M=0.2\,M_\odot$ up to $1.2\,M_\odot$.
The mass distribution exhibit both the peak at $0.562 \, M_\odot$ and the
high mass tail. Distinct asymmetry of the mass distribution and the apparent
excess of stars just below the maximum mass of $0.562 \, M_\odot$ seems a
real feature, due to the very high number of white dwarfs in Fig. 1.

The secondary peak at $1.2\, M_\odot$ obviously is not a real feature.
Due to limitations in Koester's grid of model atmospheres,
Kleinman et al. (2004) have arbitrarily assigned the maximum $\log g = 9.0$
to all white dwarfs, which probably have even higher surface gravities.
Stars with $\log g =9.0$ clump at $\approx 1.2 \, M_\odot$ in the $M-R$ 
relations for carbon -- core white dwarfs (Panei et al. 2000).

Fig. 2 presents the empirical mass-radius relation for the same set
of 1175 hot DA stars in the SDSS DR1 sample. The relation does not extend
to masses  higher than $1.2 \, M_\odot$, or rather gravities $\log g > 9.0$.
For the lowest masses we note a scatter of the points. This is because here
the surface gravities $\log g$ are much lower than in the high mass
tail of white dwarfs, hence
the thickness of atmospheres (contributing to $R$) is much larger, and is 
strongly dependent on the $\te$ of a star.  

\begin{figure}
   \resizebox{\hsize}{7.0cm}{\rotatebox{0}{\includegraphics{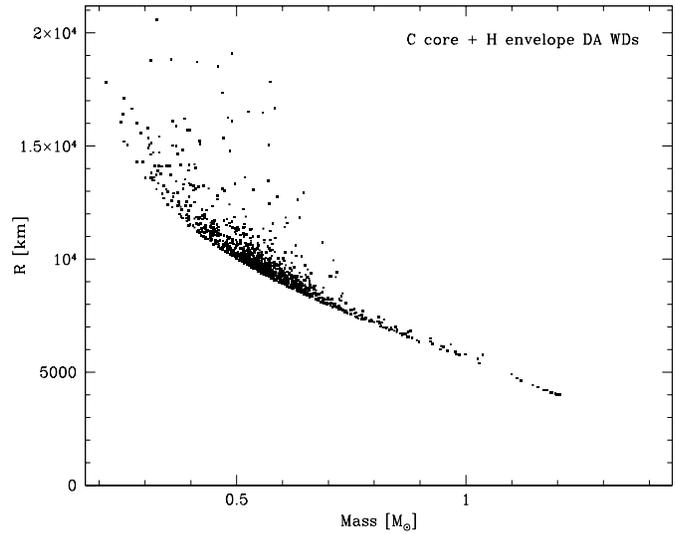}}}
   \caption[]{ Mass -- radius relation for 1175 hot white dwarfs of
      the same sample. The relation does not extend to masses $M$ 
      higher than $1.2 \, M_\odot$, or rather to the highest 
      surface gravities $\log g > 9.0$.  }
   \label{fig:fig2}
\end{figure}

\section{Theoretical M-R relations}

The mass -- radius  relations used  in this work  are those  described in
Panei  et   al.   (2000).   These  authors   have  presented  detailed
finite temperature mass -- radius relations for white dwarfs with helium,
carbon, oxygen, silicon, and iron cores based on a complete and updated
treatment of  the white dwarf evolution.  Constitutive  physics of the
stellar  models includes  the equation  of state  for  the low-density
regime from Saumon et al.  (1995) for hydrogen and helium plasmas, and
ionic,  coulomb  interactions,   partially  degenerate
electrons, electron exchange, and Thomas--Fermi contributions at finite
temperature for the high density regime.  Radiative opacities are from
OPAL  (Iglesias \& Rogers  1993) for  zero metallicity  and conductive
opacities  and neutrino  emission  rates for  helium, carbon,  oxygen,
silicon, and iron composition from Itoh and collaborators (see Althaus
\& Benvenuto 1997).  

The set of evolutionary sequences presented in
Panei  et al.   (2000) are  particularly appropriate  for  our purpose
because  they   constitute  a  homogeneous   sequence  of  mass-radius
relations.  In  particular, such  relations have been  constructed for
white dwarf masses ranging from 0.15 M$_{\odot}$ to 0.5 M$_{\odot}$ at
intervals of 5  \% for helium core composition;  from 0.45 M$_{\odot}$
to 1.2 M$_{\odot}$ at intervals of 0.01 M$_{\odot}$ for carbon, oxygen
and silicon  cores; and  from 0.45 M$_{\odot}$  to 1.0  M$_{\odot}$ at
intervals  of  0.01 M$_{\odot}$  for  the case  of  an  iron core.  

In particular, for very low mass white dwarfs we have assumed a helium core
composition because their progenitors never attain high enough central 
temperatures for helium to be ignited at their cores.  The evolutionary 
sequences were computed down to log  $L/$L$_{\odot} = -5$, and cover the 
effective temperature range from 145000 K to  4000 K. In addition, two  
cases for the hydrogen envelope have been considered: $M_{\rm H}/M= 10^{-5}$
($M_{\rm H}/M = 3 \times 10^{-4}$ for the helium core models), and
$M_{\rm H}/M = 0$. The mass of  the pure helium layer  above the core
amounts to  1~\% of the stellar mass.

\section{Mass and radius errors }

The SDSS DR1 tables list errors of their $\te$ and $\log g$ values for
each star, which we denote here as $\delta\te$ and $\delta\log g$. We 
transform them to mass and radius errors, $\delta M$ and $\delta R$, in the
following way.

Eqs. 1$-$2 imply, that small increments of all 4 parameters can be 
expressed as
\begin{equation}
\delta M = { 2R \, 10^{\log g} \over G} \, \delta R +
   { R^2 10^{\log g} \ln 10 \over G } \, \delta \log g \, ,
\label{equ:var3}
\end{equation}
\begin{equation}
\delta M = {\partial M \over \partial R} \, \delta R +
   {\partial M \over \partial \te } \, \delta \te \, .
\label{equ:var4}
\end{equation}

Eqs. \ref{equ:var3}$-$\ref{equ:var4} form the system of two algebraic
equations for two unknown numbers, $\delta M$ and $\delta R$. Errors 
$\delta\te$ and $\delta\log g$ are known from the input catalogue by
Kleinman et al. (2004), and all the coefficients (including both partial
derivatives) can easily be computed at the points ($M,R$), and the latter
were determined already. 

Solving of Eqs. \ref{equ:var3}$-$\ref{equ:var4} yields the explicit
expressions for both errors $\delta R$ and $\delta M$ (cgs units)

\begin{eqnarray}
\delta R &=& \left({R^2 \over G} 10^{\log g} \ln 10 \> \delta \log g -
   {\partial M \over \partial \te} \, \delta \te \right)  \\
   & & \hskip12mm \times \hskip2mm \left({{\partial M \over {\partial R}}
   - {2R \over G} 10^{\log g} } \right) ^{-1}   \, ,  \nonumber
\label{equ:var5}
\end{eqnarray}
\begin{equation}
\delta M = { R^2 \over G} 10^{\log g} \ln 10 \> \, \delta \log g +
   { 2R \over G } 10^{\log g} \, \delta R \, .
\label{equ:var6}
\end{equation}
Finally, we accepted absolute values of the above expressions as the real
$\delta M$ and $\delta R$ errors. 

The average mass and radius errors resulting from our analysis of the
SDSS DR1 catalogue of DA white dwarfs are: $\langle\delta M\rangle=0.059
M_\odot$, and $\langle \delta R \rangle =792 $ km, respectively.

\section{Core composition of white dwarf stars }

White dwarfs are thought to have a carbon/oxygen core, the exact proportions
of which are uncertain and fixed by the mixing processes occurred prior to
the formation of a white dwarf. Unfortunately, theoretical $M-R$ relations 
for white dwarfs having cores with mixtures of carbon and oxygen were not
available at the time when our research was performed. However, we believe
that the pure carbon and pure oxygen core compositions of Panei et al. (2000)
would bracket the real one. Also, for stellar masses larger than
1.05 solar masses, the core composition is probably a mixture
of oxygen and neon, being oxygen the dominant one, and for $M$ lower
than $0.45 M_\odot$ a pure helium core is expected.

\begin{figure}
   \resizebox{\hsize}{7.0cm}{\rotatebox{0}{\includegraphics{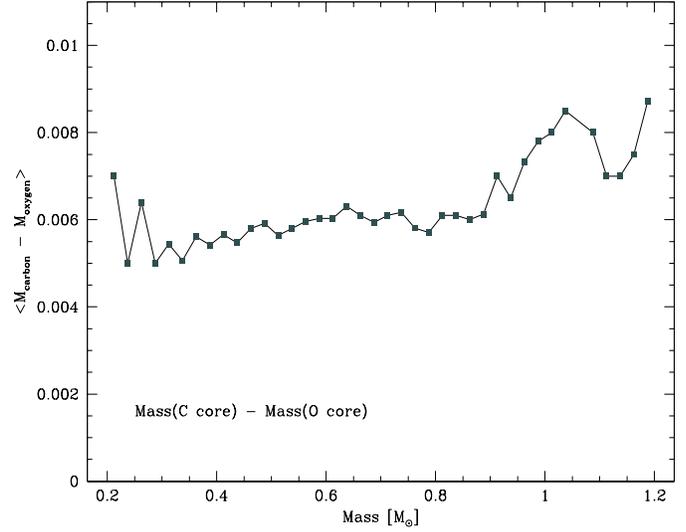}}}
   \caption[]{The difference between DA white dwarf masses obtained from 
      $M-R$ relations corresponding either to pure carbon, or pure oxygen
      cores. All the 1175 stars were grouped in bins of the width $0.025
      M_\odot$. Dots correspond to the average difference $M({\rm C} \>\> 
      {\rm core}) - M({\rm O} \>\> {\rm core})$ in each group.       }
   \label{fig:fig3}
\end{figure}

We have determined a similar set of white dwarf masses and radii assuming
$M-R$ relations corresponding to the pure oxygen core plus H envelope of the
same mass as for the pure carbon core, $M_{\rm H}/M = 10^{-5}$. Resulting
differences between both grids of masses are very small. The mass of a carbon
core DA white dwarf was always higher than the oxygen core mass of the same
star in the whole range of $0.2 - 1.2 \, M_\odot$. The average difference 
between masses of both types is equal to $0.00592\, M_\odot$ in our sample of
1175 stars, see Fig. 3.

All the stars in our sample were grouped in bins of the width $0.025
\, M_\odot$, similarly as in Fig. 1. Dots in Fig. 3 correspond to the 
average difference $M({\rm C} \>\> {\rm core})-M({\rm O} \>\> {\rm core})$
in each bin. One can note, that there is little evidence of a systematic
trend in the run of the difference $M({\rm C} \>\> {\rm core})-M({\rm O}
\>\> {\rm core})$ vs. mass $M({\rm C} \>\> {\rm core})$. The above 
difference increases with a white dwarf mass only insignificantly.

Similar comparison was also performed for DA white dwarfs with carbon
masses $M \le 0.45 \, M_{\odot}$. We have found 71 such stars in our
sample of 1175,
and computed the differences $M({\rm He} \>\> {\rm core})-M({\rm C} \>\> 
{\rm core})$ between masses derived either from the $M-R$ corresponding 
either to pure helium, or pure carbon cores , respectively. Spectroscopic mass
of a helium core white dwarf was always higher than the mass of a carbon
core star, and the average difference amounts to $0.0376 \, M_{\odot}$.

\section{Concluding remarks}

\noindent {\bf 1.}
We have determined the distribution of DA white dwarf masses for the
magnitude limited sample of 1175 stars in the SDSS DR1 table. Our mass
distribution exhibits the peak at $M=0.562 \, M_\odot$, and slight
asymmetry towards both higher and lower masses. 

The peak mass of this paper is in excellent agreement with earlier studies,
which used much less numerous samples and various methods of white dwarf mass
determination. We present here a brief listing of previously determined
peak masses of DA white dwarfs. Unfortunately, the early paper by
Shipman (1979) does not provide data corresponding to data of
other papers, which are collected in Table 1.

There exists a striking consistency of all peak mass determinations, in
spite of relatively nonnumerous samples in the above listing. Our research
adds the peak mass $M_{\rm peak} = 0.562 \, M_\odot$ to the list
(carbon core stars),
which perfectly confirms the value of $M_{\rm peak}$ by Bergeron, Saffer
\& Liebert (1992). Moreover, we claim that have reliably determined the
shape of mass distribution of DA white dwarfs, except for the single
bin in Fig. 1, corresponding to the most massive stars.

\vskip2mm
\noindent {\bf 2.}
We note, that spectroscopic masses $M$ and radii $R$ in the present and
earlier papers can still exhibit systematic errors due to
uncertainties in the present theories of plasma physics. Theoretical 
profiles of the Balmer lines in spectra of DA white dwarfs are very strongly
broadened, and the dominant contributor is the pressure broadening of line
wings. The correct description of the pressure broadening in dense plasma
of high gravity white dwarf atmospheres is a very difficult problem.

Recently Wujec et al. (2002) proposed the new method of Stark 
broadening calculations based on simulation techniques. Their theory
is able to compute the broadening, shift, and asymmetry parameters of
Stark broadened hydrogen line opacity profile. Previous Stark broadening
calculations were not able to include all the three plasma effects
simultaneously. However, no grids of theoretical Balmer line profiles in
white dwarf spectra by Wujec et al. (2002) are currently available.

\vskip2mm
\noindent {\bf 3.}
Catalogues of masses, radii, resulting bolometric luminosities for both 
C -- core and O -- core white dwarfs,
and all the input parameters are available from authors on request.
Bolometric luminosities and their errors are given by the following
expressions
\begin{equation}
L_{\rm bol} = 4\pi R^2 \, \sigma_{\rm rad} \te ^4 \, ,
\end{equation}
\begin{equation}
\delta L_{\rm bol} = 8\pi R \,\sigma_{\rm rad}\te^3 \, \left(\te \,\delta R 
   + 2 R \, \delta\te \right) \, .
\end{equation}

\begin{table}
\caption{Peak masses of the published DA white dwarf mass distributions, 
   in  $M_{\rm peak} / M_\odot $.  }
\renewcommand{\arraystretch}{1.1}
\begin{tabular}{rll}
-- & $0.58 $ & Koester, Schulz \& Weidemann (1979), average \\
   &         & mass of 122 DA white dwarfs                  \\
-- & $0.603$ & Weidemann \& Koester (1984), 70 DA WD stars     \\
-- & $0.571$ & McMahan (1989), 53 DA white dwarfs              \\
-- & $0.60 $ & Weidemann (1990)                                \\
-- & $0.562$ & Bergeron, Saffer \& Liebert (1992), 129 DA white\\
   &         & dwarfs                                          \\
-- & $0.56 $ & Liebert \& Bergeron (1995), 200 white dwarfs from \\
   &         & the Palomar Green survey (Green et al. 1986)   \\
-- & $0.570$ & Finley, Koester \& Basri (1997), 174 DA white   \\
   &         & dwarfs, some with cool companions               \\
-- & $0.55 $ & Marsh et al. (1997a,b), 89 stars of the ROSAT  \\
   &         & All-Sky X-ray and EUV Surveys                  \\
-- & $0.56 $ & Vennes et al. (1997), 110 EUV selected DA WDs   \\
-- & $0.57 $ & Vennes (1999), 141 EUV/X-ray selected WDs       \\
\end{tabular}
\end{table}

\begin{acknowledgements}

JM and MN acknowledge support by grant No. 1 P03D 001 26 from
the Polish Committee for Scientific Research.
LGA acknowledges the Spanish MCYT for a Ram\'on y Cajal Fellowship.
We are grateful to the referee for his remarks on the original 
version of our paper.

% Non-commercial scientific and technical publications using SDSS data 
% should include the following acknowledgment:

Funding for the creation and distribution of the SDSS Archive has been
provided by the Alfred P. Sloan Foundation, the Participating Institutions,
the National Aeronautics and Space Administration, the National Science 
Foundation, the U.S. Department of Energy, the Japanese Monbukagakusho, 
and the Max Planck Society. The SDSS Web site is
http://www.sdss.org/.

The SDSS is managed by the Astrophysical Research Consortium (ARC) for the
Participating Institutions. The Participating Institutions are The
University of Chicago, Fermilab, the Institute for Advanced Study, the Japan
Participation Group, The Johns Hopkins University, Los Alamos National
Laboratory, the Max-Planck-Institute for Astronomy (MPIA), the
Max-Planck-Institute for Astrophysics (MPA), New Mexico State University,
University of Pittsburgh, Princeton University, the United States Naval
Observatory, and the University of Washington.

\end{acknowledgements}

\end{document}